\documentclass[pre,twocolumn,showpacs]{revtex4-1}
\usepackage{amsmath,amssymb}
\usepackage{graphicx,color}
\usepackage{times}
\usepackage[pdftex,bookmarks,colorlinks,breaklinks]{hyperref}
\makeatletter
\hypersetup{linkcolor=blue,citecolor=blue,filecolor=dullmagenta,urlcolor=blue, 
 pdftitle={Fluctuations and correlations in scattering on a resonance coupled to a chaotic background},
 pdfauthor={Dmitry V. Savin}, pdflang={English} }

\newcommand{\aver}[1]{\left \langle #1 \right \rangle}

\newcommand{\Hbg}{H_{\mathrm{bg}}}
\newcommand{\Rbg}{R_{\mathrm{bg}}}
\newcommand{\phasebg}{\theta_{\mathrm{bg}}}
\newcommand{\Szero}{S^{(0)}}
\makeatother

\begin{document}

\title{Fluctuations and correlations in scattering on a resonance coupled to a chaotic background}
\author{D. V. Savin}
 \affiliation{Department of Mathematics, Brunel University London, Uxbridge, UB8 3PH, United Kingdom}
\published{in \emph{Acta Phys. Pol. A \textbf{132}, 1688 (2017)}\,}

\begin{abstract}
We discuss and briefly overview recent progress with studying fluctuations in scattering on a resonance state coupled to the background of many chaotic states. Such a problem arises naturally, e.g., when dealing with wave propagation in the presence of a complex environment. Using a statistical model based on random matrix theory, we obtain a number of nonperturbative results for various statistics of scattering characteristics. This includes the joint and marginal distributions of the reflection and transmission intensities and phases, which are derived exactly at arbitrary coupling to the background with finite absorption. The intensities and phases are found to exhibit highly non-trivial statistical correlations, which remain essential even in the limit of strong absorption. In the latter case, we also consider the relevant approximations and their accuracy. As an application, we briefly discuss the statistics of the phase rigidity (or mode complexness) in such a scattering situation.
\end{abstract}
\pacs{05.45.Mt, 03.65.Nk, 05.60.Gg, 24.60.-k}
\maketitle
\setlength{\unitlength}{1cm}
\begin{picture}(0,0)(-1.575,-0.65)
 \put(0,0){\parbox{0.8\textwidth}{\footnotesize Proceedings of 8th Workshop on Quantum Chaos and Localisation Phenomena, Warsaw, Poland: May 19--21, 2017}}
\end{picture}
\vspace*{-5ex}

\subsection{Introduction and formalism}\label{sec:intro}

Scattering on a well isolated resonance is a textbook example giving rise to the natural (``bell-shaped'') spectral profile of the transmission intensity \cite{Mahaux,Nussenzveig,Mello,savi16}. In the vicinity of the resonance with energy $\varepsilon_0$ and width $\Gamma_0$, the integral contribution of all other (remote) possible resonances amounts to scattering phases with a weak energy dependence. In a resonance approximation, one usually neglects such a non-resonant contribution (which can also be eliminated by other means \cite{nish85}). The scattering amplitude between two open channels $a$ and $b$ is then given by a multichannel Breit-Wigner formula \cite{Mahaux}
\begin{equation}\label{S0_ab}
  \Szero_{ab}(E) = \delta_{ab} - i \frac{A_a A_b }{ E-\varepsilon_0+\frac{i}{2}\Gamma_0}\,.
\end{equation}
For simplicity, we assume here invariance under time reversal, so $S$ is a symmetric matrix with the real channel amplitudes $A_c$. The latter determine the partial (per channel) decay widths $A^2_{c}$ and the total escape width $\Gamma_0=\sum_{c}A^2_c$, with the sum running over all channels open at the given scattering energy $E$. This ensures the unitarity of the $S$ matrix (at real $E$).

In many situations, such a resonance represents a specific ``simple'' (deterministic) excitation that is coupled to the background of many ``complicated'' (usually chaotic) states \cite{soko97}. The prominent examples include giant resonances and doorway states in atomic and nuclear physics \cite{Bohr,harn86,soko97i,gu99,zele16} and similar mechanisms in open mesoscopic systems \cite{aber08,guhr09,soko10,mora12}. Because of this coupling the initial amplitude is spread over the background with a rate determined by the so-called spreading width $\Gamma_{\downarrow}$ \cite{soko97,Bohr}. This gives rise to an effective damping, resulting in the mean (``optical'') $S$ matrix
\begin{equation}\label{S_ab}
  \overline{S_{ab}}(E) = \delta_{ab} - i \frac{A_a A_b }{ E-\varepsilon_0+\frac{i}{2}(\Gamma_0+\Gamma_{\downarrow})}\,,
\end{equation}
where the average is performed over the fine energy structure of the complex background.   The arising competition between two decay mechanisms leads to a strong suppression of transmission through such a simple state when the ratio $\eta\equiv\Gamma_{\downarrow}/\Gamma_0$ of the spreading ($\Gamma_{\downarrow}$) to escape ($\Gamma_0$) width exceeds unity \cite{savi17}.

In the context of wave transport, the model provides a useful approach for quantifying fluctuations in an established transmission signal induced by a complex environment \cite{savi17}. The applicability of the model is actually much broader \cite{savi17a}. It covers the whole range of the scattering observables,  including joint statistics of the intensities and phases \cite{savi17b}.

The approach developed in \cite{savi17,savi17a,savi17b} enables us to fully describe the influence of the chaotic background with finite dissipative losses on the resonance scattering. In short, it makes use of the well-known strength function formalism \cite{soko97,Bohr} to account for coupling to the background and employs random matrix theory (RMT) \cite{guhr98,Stoeckmann,fyod11ox} to model the chaotic nature of the latter. We are interested in fluctuations in scattering at the resonance energy $\varepsilon_0$, corresponding to the peak of the original signal. The starting point is the following convenient representation of the $S$ matrix at $E=\varepsilon_0$ derived in \cite{savi17,savi17a}:
\begin{equation}\label{S}
  S = 1 - \frac{1}{1+i\eta K}(1-S^{(0)})\,.
\end{equation}
Here $\Szero$ stands for the $S$ matrix (\ref{S0_ab}) in the ``clean'' system without the background. $K$ denotes the local Green's function \cite{fyod04b} of the background defined by $K = \frac{N\Delta}{\pi}(\frac{1}{\varepsilon_0-\Hbg})_{11}$,  which is a (rescaled) diagonal element of the resolvent of the Hamiltonian $\Hbg$ describing $N\gg1$ background states with the mean level spacing $\Delta\sim\frac{1}{N}$. In the chosen units, the constant $\eta$ becomes the only parameter controlling the strength of coupling to the background. Finite absorption is then taken into account by uniform broadening $\Gamma_\mathrm{abs}$ of the background states, resulting in complex $K$ \cite{fyod04b}
\begin{equation}\label{K}
  K =  u - iv, \quad v>0,
\end{equation}
with the normalisation $\aver{iK}=\aver{v}=1$. Within the RMT approach, $\Hbg$ can be modelled by the Gaussian orthogonal ensemble (GOE) of random matrices, which enables us to fully characterise the universal statistics of $K$ at arbitrary absorption rate $\gamma\equiv2\pi\Gamma_\mathrm{abs}/\Delta$ \cite{fyod04b,savi05,fyod05}. Importantly, $u$ and $v$ are the mutually correlated random variables that have the following joint probability density function (jpdf) \cite{fyod04b}:
\begin{align} \label{P(u,v)}
  \mathcal{P}(u,v) = \frac{1}{2\pi v^2}P_0(x), \quad x=\frac{u^2+v^2+1}{2v}>1.
\end{align}
The function $P_0(x)$ has the meaning of a probability distribution and its explicit form is known exactly at any $\gamma$ \cite{savi05,fyod05}.

We now use these results to first discuss in Sec.~\ref{sec:ampl} the arising scattering pattern and how the two control parameters $\eta$ and $\gamma$ can be extracted from the scattering data.  Sections~\ref{sec:jpfd1} and ~\ref{sec:jpfd2} then provide the detailed analysis of the statistical properties of the transmission and reflection amplitudes in terms of various joint distributions. An application of the obtained results to the so-called phase rigidity is considered in Sec.~\ref{sec:phase}. Finally, we conclude with a brief summary in Sec.~\ref{sec:summary}.

\subsection{Transmission and reflection amplitudes}\label{sec:ampl}

Let us consider a two-channel setup, which is sufficient for the discussion. The $S$ matrix can then be parameterised as
\begin{equation}\label{S_t,r}
  S = \left(\begin{matrix}
        r_+ & t \\  t & r_-
      \end{matrix} \right)
\end{equation}
in term of the transmission ($t$) and two reflection amplitudes ($r_\pm$). Their explicit forms are found from (\ref{S}) as follows
\begin{equation}\label{t,r_K}
 \begin{array}{l}
  t = t_0 / (1+\eta v +i\eta u) \equiv\sqrt{T} e^{i\theta_T}, \\[1ex]
  r_{\pm} = (\eta v \pm r_0+i\eta u)/(1+\eta v +i\eta u) \equiv\sqrt{R_{\pm}} e^{i\theta_R^{\pm}}.
 \end{array}
\end{equation}
The amplitudes $t_0$ and $r_0$ describe the direct coupling between the channels avoiding the background ($\eta=0$). They satisfy the flux conservation, $t_0^2+r_0^2=1$. The flux is no longer conserved in scattering with the background at finite absorption, when $S$ becomes subunitary. Such a unitarity deficit can be naturally quantified by the following matrix \cite{savi03a}:
\begin{equation}\label{D}
  1-S^\dagger S = (1-S_0) d,
  \quad d \equiv \frac{2\eta v}{(1 + \eta v)^2+\eta^2u^2}.
\end{equation}
The positive quantity $d\leq\frac{1}{2}$ is therefore a useful measure of the total losses dissipated in the background \cite{savi17a}.

At vanishing absorption, $\gamma=0$, we have $v=0$ (thus $d=0$) identically. The intensities and phases are then deterministic functions of each other defined by the following relations:
\begin{equation}\label{T,R_gam0}
 \begin{array}{l}
  T = t_0^2\cos^2\theta_T = 1-R_{\pm}, \\[1ex]
  \theta_R^{\pm} = \frac{\pi}{2} + \theta_T \pm \arctan(r_0\cot\theta_T),
 \end{array}
\end{equation}
where $\theta_T=-\arctan(\eta u)$. Therefore, their distributions are determined by the random variable $u$ that  is known \cite{fyod04b,mell95} to be  Cauchy distributed in this limit. This yields, in particular, the following distribution of the transmission phase \cite{savi17b}:
\begin{equation}\label{Pphase0}
  \mathcal{P}_{0}(\theta_T) = \frac{1}{\pi (\eta\cos^2\theta_T+\eta^{-1}\sin^2\theta_T)},
\end{equation}
which provides all others by a suitable change of variables.

At finite absorption, the correlations imposed by the above (flux conservation) constraint (\ref{T,R_gam0}) are removed. We find
\begin{equation}\label{T,R}
 \begin{array}{l}
  T = t_0^2/[(1+\eta v)^2+\eta^2u^2]\,,  \\[1ex]
  R_{\pm} = [(\eta v \pm r_0)^2+\eta^2u^2]/[(1 + \eta v)^2+\eta^2u^2]
\end{array}
\end{equation}
for the transmission and reflection coefficients, and
\begin{equation}\label{theta_TR}
 \begin{array}{l}
  \tan\theta_T = -\eta u/(1+\eta v)\,,  \\[1ex]
  \tan\theta_R^{\pm} =(1 \mp r_0)\eta u/[(1+\eta v)(\eta v \pm r_0)+\eta^2u^2]\
\end{array}
\end{equation}
for the corresponding phases. Thus they have non-trivial joint statistics determined by the distribution (\ref{P(u,v)}) of $u$ and $v$ \cite{savi17a,savi17b}.

It is instructive to complement the above description of the scattering pattern by the ``external'' viewpoint in terms of the interference between the deterministic scattering phase (due to the direct transmission) and the random phase induced by the background \cite{savi17}.  To this end, we note that both $S$ and $S_0$ can be diagonalised by an orthogonal matrix $O_{\varphi}$, where angle $\varphi$ expresses the degree of channel nonorthogonality. Then
\begin{equation}\label{S_phases}
  S = O_{\varphi} \left(\begin{matrix}
        1 & 0 \\  0 & -S_\mathrm{bg}
      \end{matrix} \right)O_{\varphi}^T,
\end{equation}
where $S_\mathrm{bg}=\frac{1-i\eta K}{1+i\eta K}\equiv\sqrt{\Rbg}e^{i\phasebg}$ stands for the background contribution into the full scattering process. This yields an alternative parametrisation of the scattering amplitudes
\begin{equation}\label{t,r_bg}
 \begin{array}{l}
  t = \frac{\sin2\varphi}{2}(1+\sqrt{\Rbg}e^{i\phasebg}), \\[1ex]
  r_{\pm} = \frac{1}{2}(1-\sqrt{\Rbg}e^{i\phasebg}) \pm \frac{\cos2\varphi}{2}(1+\sqrt{\Rbg}e^{i\phasebg}).
 \end{array}
\end{equation}
Note that the background acts as a single-channel scattering centre. For such a situation, the joint distribution of the reflection coefficient $\Rbg$ and phase $\phasebg$ is also known exactly \cite{savi05,fyod05} at arbitrary absorption (see \cite{kuhl05} for the relevant experimental study). In particular, we have  $\aver{S_\mathrm{bg}}=\frac{1-\eta}{1+\eta}$, yielding the average transmission and reflection amplitudes as follows
\begin{equation}\label{t,r_mean}
  \aver{t}=\frac{\sin2\varphi}{1+\eta}, \quad \aver{r_{\pm}}=\frac{\eta\pm\cos2\varphi}{1+\eta}.
\end{equation}

These expressions are useful for determining the model parameters when applying the approach to real systems. By measuring the average amplitudes and fitting them to (\ref{t,r_mean}), one can extract the background coupling $\eta$ and the mixing phase $\varphi$, thus fixing $t_0=\sin2\varphi$ and $r_0=\cos2\varphi$. (Note that both $\aver{t}$ and $\aver{r_{\pm}}$ are real, which should be helpful in eliminating global phases that might be present in realistic situations.) The absorption rate $\gamma$ can then be obtained independently from the correlation analysis of scattering spectra \cite{schaf03,kuhl05a}. This allows one to perform comparison with the experimental data without fit parameters. We now discuss theoretical predictions for joint statistics of both intensities and phases in more detail.

\subsection{Joint distribution of the intensities}\label{sec:jpfd1}

\subsubsection{Perfect coupling (no backscattering)}
\begin{figure}
  \centering
  \includegraphics[width=.985\linewidth]{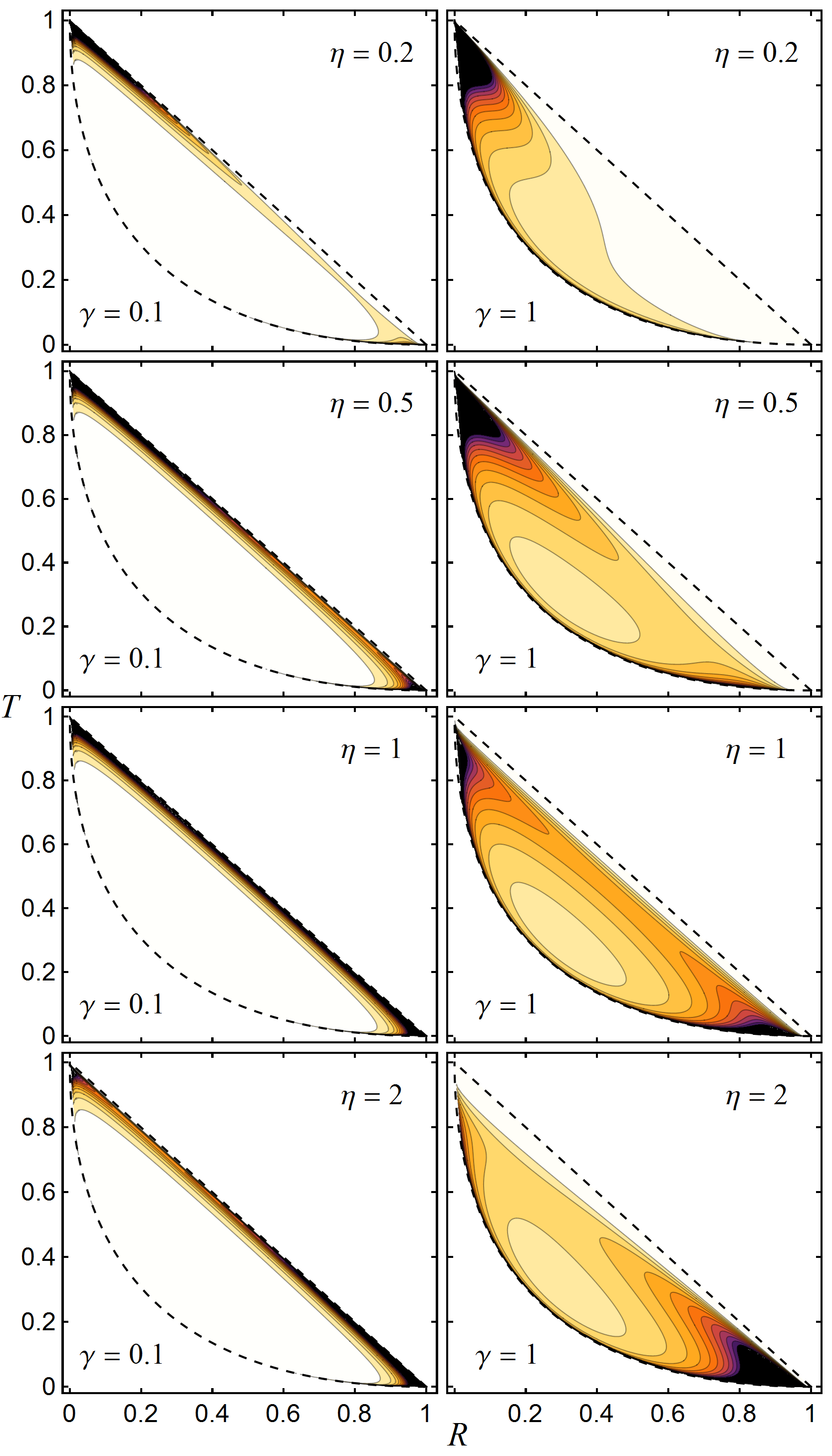}
  \caption{Contour plots of the joint distribution (\ref{P(r,t)}) of the reflection and transmission intensities at perfect coupling. The absorption rate $\gamma=0.1$ (left) or $1$ (right) and the background coupling $\eta$ is increased as $\eta=0.2, 0.5,1,2$ (from top to bottom). Darker regions correspond to higher values of the jpdf. Dashed lines indicate the boundaries of the distribution support. Note the symmetry of the distribution at $\eta=1$ and also at the reciprocal values of $\eta$.
  \label{fig:jpdf_rt} }
\end{figure}

We consider first the case of perfect coupling $t_0=1$ ($r_0=0$). Making use of relations (\ref{T,R}) and applying the Jacobian calculus, the joint distribution $\mathcal{P}(R,T)$ of the reflection and transmission intensities is expressed in terms of the known function $P_0$ by the following attractive formula \cite{savi17a}:
\begin{equation}\label{P(r,t)}
  \mathcal{P}(R,T) = \frac{2}{\pi(1-R-T)^2\sqrt{y}} P_0\biggl(\frac{\eta^{-1}R+\eta T}{1-R-T}\biggr)\,,
\end{equation}
being nonzero only in the region defined by $1-R-T>0$ and $y=1+2RT-(1-R)^2-(1-T)^2>0$. It follows at once that function (\ref{P(r,t)}) has the following important symmetry
\begin{equation}\label{sym_joint}
  \mathcal{P}(R,T)|_{\eta} = \mathcal{P}(T,R)|_{\eta^{-1}}
\end{equation}
under the interchange $\eta\to\eta^{-1}$. The symmetry property (\ref{sym_joint}) holds at arbitrary absorption. This shows that the coupling strength $\eta$ controls the weight of the total flux distribution between its reflection and transmission sectors. In particular, distribution (\ref{P(r,t)}) becomes symmetric with respect to the line $R=T$ at the special coupling $\eta=1$ (see further Fig.~\ref{fig:jpdf_rt}).

 In the limit of vanishing absorption, $\gamma\to0$, one can use \cite{fyod04b} that $P_0(x)\to\delta(\frac{1}{x})$. This yields the following result:
\begin{equation}\label{P(r,t)gam0}
  \mathcal{P}_{\gamma=0}(R,T) =  \delta(1-R-T) \mathcal{P}_{0}(T).
\end{equation}
The first (singular) factor here stands for the conditional pdf of $R$, expressing in the present context the flux conservation (\ref{T,R_gam0}). The marginal distribution $\mathcal{P}_0(T)$ of $T$ is given by \cite{savi17}
\begin{align}\label{PT_stable}
  \mathcal{P}_{0}(T)= \frac{1}{\pi\sqrt{T(1-T)}}\frac{1}{\eta T + \eta^{-1}(1-T)}\,.
\end{align}
As discussed above, this form follows from the Cauchy distribution of the random variable $u$ in this limit.

At finite absorption, the function $P_0(x)$ gets exponentially suppressed ($\sim e^{-\gamma x/4}$) for large $x\gg1$. As a result, the jpdf $\mathcal{P}(R,T)$ at small $\gamma$ is mostly concentrated within a thin layer $\sim\gamma\ll1$ near the boundary $R+T=1$.  When $\gamma$ is increased, the distribution starts exploring its whole support. The typical behavior of the distribution is illustrated on Fig.~\ref{fig:jpdf_rt} displaying the density plots at various values of $\eta$ and $\gamma$. It clearly shows  highly non-trivial correlations between reflection and transmission. Note, however, the symmetry (\ref{sym_joint}) of the distribution at the reciprocal values of $\eta$, which holds at any $\gamma$.

\begin{figure}
  \centering
  \includegraphics[width=.925\linewidth]{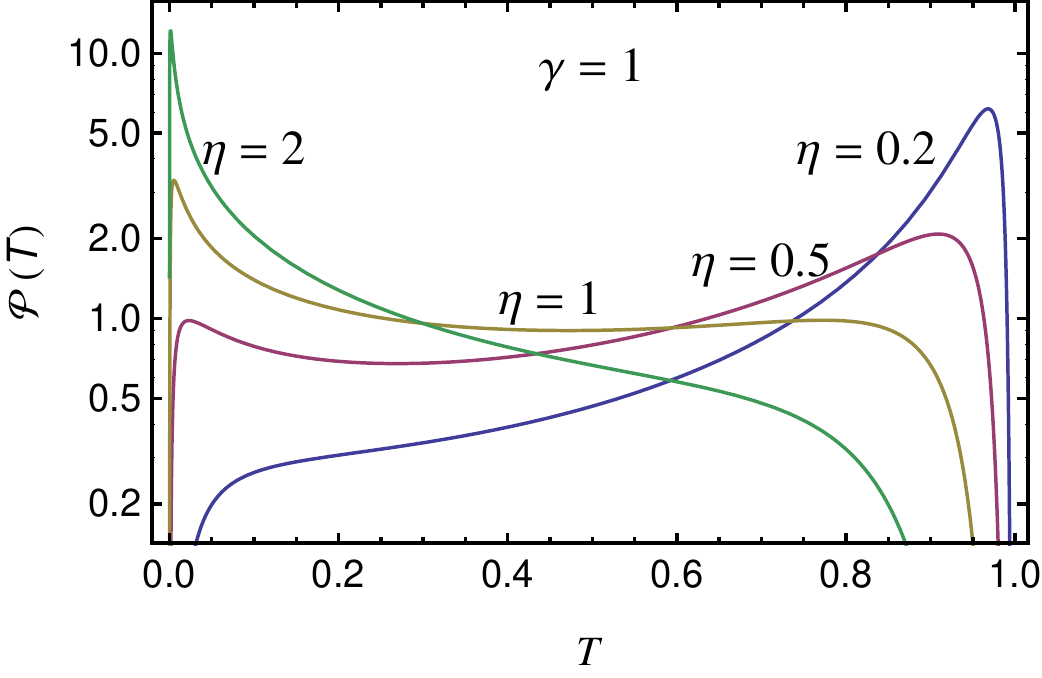}
  \caption{Transmission distribution (\ref{Ptr}) at perfect coupling for $\gamma=1$ and the same values of $\eta$ as in Fig.~\ref{fig:jpdf_rt}. By the symmetry property (\ref{Pref}), the corresponding reflection distributions would be given by the same curves at the reciprocal values of $\eta$.
  \label{fig:p(t)} }
\end{figure}

The marginal distributions can now be obtained by integrating (\ref{P(r,t)}) over $R$ or $T$. One finds the following expression for the transmission distribution ($0\le{T}\le1$) \cite{savi17a}:
\begin{equation}\label{Ptr}
  \mathcal{P}(T) = \int_{\rho_-}^{1-T}\!\!\frac{dR}{\pi(1-R-T)^2}
  \frac{2P_0\bigl(\frac{\eta^{-1}R+\eta T}{1-R-T}\bigr) }{ \sqrt{(\rho_{+}-R)(R-\rho_{-})} }\,,
\end{equation}
with $\rho_{\pm}=(1\pm\sqrt{T})^2$. One can further show that this expression is equivalent to the one derived recently in \cite{savi17}. The advantage of representation (\ref{Ptr}) is that it utilizes the symmetry property (\ref{sym_joint}) explicitly. In particular, the distribution of reflection is simply related to that of transmission as follows
\begin{equation}\label{Pref}
  \mathcal{P}^{\mathrm{(ref)}}(R)|_{\eta} = \mathcal{P}^{\mathrm{(tr)}}(R)|_{\eta^{-1}}\,.
\end{equation}
This remarkable relation shows that despite lacking any apparent connection between the reflection and transmission coefficients at finite absorption, their distribution functions turn out to be linked by symmetry (\ref{Pref}). With explicit formulae for $P_0$ found in \cite{savi05,fyod05}, Eqs.~(\ref{P(r,t)}), (\ref{Ptr}) and (\ref{Pref}) provide the exact solution to the problem at arbitrary $\eta$ and $\gamma$.

It is possible to perform further analysis in the physically interesting limiting cases of weak and strong  absorption, when the function $P_0$ is known \cite{fyod04b} to have simple asymptotics. At $\gamma\ll1$, one has
$P_0(x)\approx\frac{2}{\sqrt{\pi}} (\frac{\gamma}{4})^\frac{3}{2} \sqrt{x+1} e^{-\frac{\gamma}{4}(x+1)}$,
yielding the leading-order correction factor $\exp[-\frac{\gamma}{8\eta}\frac{(1+(\eta-1)\sqrt{T})^2}{\sqrt{T}(1-\sqrt{T})}]$ to the zero-absorption distribution (\ref{PT_stable}). Thus, the bulk of distribution (\ref{Ptr}) is essentially reproduced by $\mathcal{P}_0(T)$ in this limit. The correction factor becomes crucial near the edges, where the exact distribution has an exponential cutoff. In the opposite case of $\gamma\gg1$, making use of $P_0(x)\approx\frac{\gamma}{4} e^{-\frac{\gamma}{4}(x-1)}$ results in the following approximation at strong absorption \cite{savi17}:
\begin{equation}\label{Pgam_strong}
    \mathcal{P}_{\gamma\gg1}(T)\approx  \frac{ \sqrt{\gamma\eta}
      \exp\left[-\frac{\gamma}{8\eta}\frac{(1-(\eta+1)\sqrt{T})^2}{\sqrt{T}(1-\sqrt{T})}\right]
     }{ 4\sqrt{\pi}(1-\sqrt{T})T^{3/4}\sqrt{1+(\eta^2-1)T} }.
\end{equation}
Figure~\ref{fig:p(t)} shows the exact distribution (\ref{Ptr}) at moderate $\gamma=1$.

\subsubsection{Nonperfect coupling}

In the general case of nonperfect coupling, $r_0\neq0$, only a part (given by $t_0^2=1-r_0^2$) of the incoming flux contributes to the transmission. In view of (\ref{T,R}), the transmission distribution is then obtained by a simple rescaling of expression (\ref{Ptr}). The reflection distribution takes a more elaborate form because of the interference between the two reflected waves, the one backscattered directly at the channel interface and the one originating from the background.  By expressing $R_{\pm}$ in terms of the variables $T$ and $R$ at $r_0=0$ studied above, one can derive the corresponding distributions in the closed form. The distribution of $R=R_+$ (and similarly for $R_-$ by changing $r_0\to-r_0$ below) reads as follows \cite{savi17a}
\begin{equation}\label{P(R)}
  \mathcal{P}(R) = \int_{T_-}^{T_*}\!\! \frac{dT}{\pi(1{-}R{-}T)^2}\frac{2P_0(x)}{\sqrt{(T_{+}-T)(T-T_{-})}}\,,
\end{equation}
where $T_*=\min(1-R,T_+)$,  $T_{\pm}=\frac{1+r_0}{1-r_0}(1\pm\sqrt{R})^2$, and
\begin{equation}\label{x}
  x = \frac{(1+r_0)(R-r_0)+T(\eta^2+r_0)}{\eta(1+r_0)(1-R-T)}\,.
\end{equation}
It reduces to Eq.~(\ref{Pref}) at perfect coupling, $r_0=0$.

A particular feature of the reflection distribution (\ref{P(R)}) is the dependence of its support on the sign of $r_0$ (see also \cite{savi17}). The distribution vanishes identically for $R\leq1-T_0$, when $r_0>0$, and covers the whole range $0\leq R\leq1$, when $r_0<0$. This follows from the compatibility requirement $T_-<T_*$ and is, of course, in agreement with definition (\ref{T,R}).

As a particular application of the above results, one can study the statistics of total losses in the system. The distribution of the unitarity deficit $d=1-R-T$ can be found in an exact form, see Ref.~\cite{savi17a} for further discussion.

\subsection{Joint intensity-phase distribution}\label{sec:jpfd2}

The joint distribution $\mathcal{P}(T,\theta)$ of the transmission intensity and phase $\theta\equiv\theta_T$ can be derived and studied along the similar lines. (One can set $t_0=1$ throughout without loss of generality.) We find the following exact representation \cite{savi17b}:
\begin{equation}\label{Pjoint}
  \mathcal{P}(T,\theta) =
  \frac{\Theta(\cos\theta-\sqrt{T})}{4\pi T(\cos\theta-\sqrt{T})^2}
  P_0[\mathrm{x}_\eta(T,\theta)],
\end{equation}
where $\Theta(x)$ is the Heaviside step function and
\begin{equation}\label{x}
 \mathrm{x}_\eta(T,\theta) =
 \frac{T(1+\eta^2)-2\sqrt{T}\cos\theta+1}{2\eta\sqrt{T}(\cos\theta-\sqrt{T})}.
\end{equation}
The joint distribution (\ref{Pjoint}) is nonzero for $0\leq T \leq\cos^2\theta$. Its profile within this region is controlled by two parameters $\gamma$ and $\eta$. Therefore, the transmission intensity and phase exhibit strong statistical correlations at finite absorption.

\begin{figure}[b]
  \centering
  \includegraphics[width=.985\linewidth]{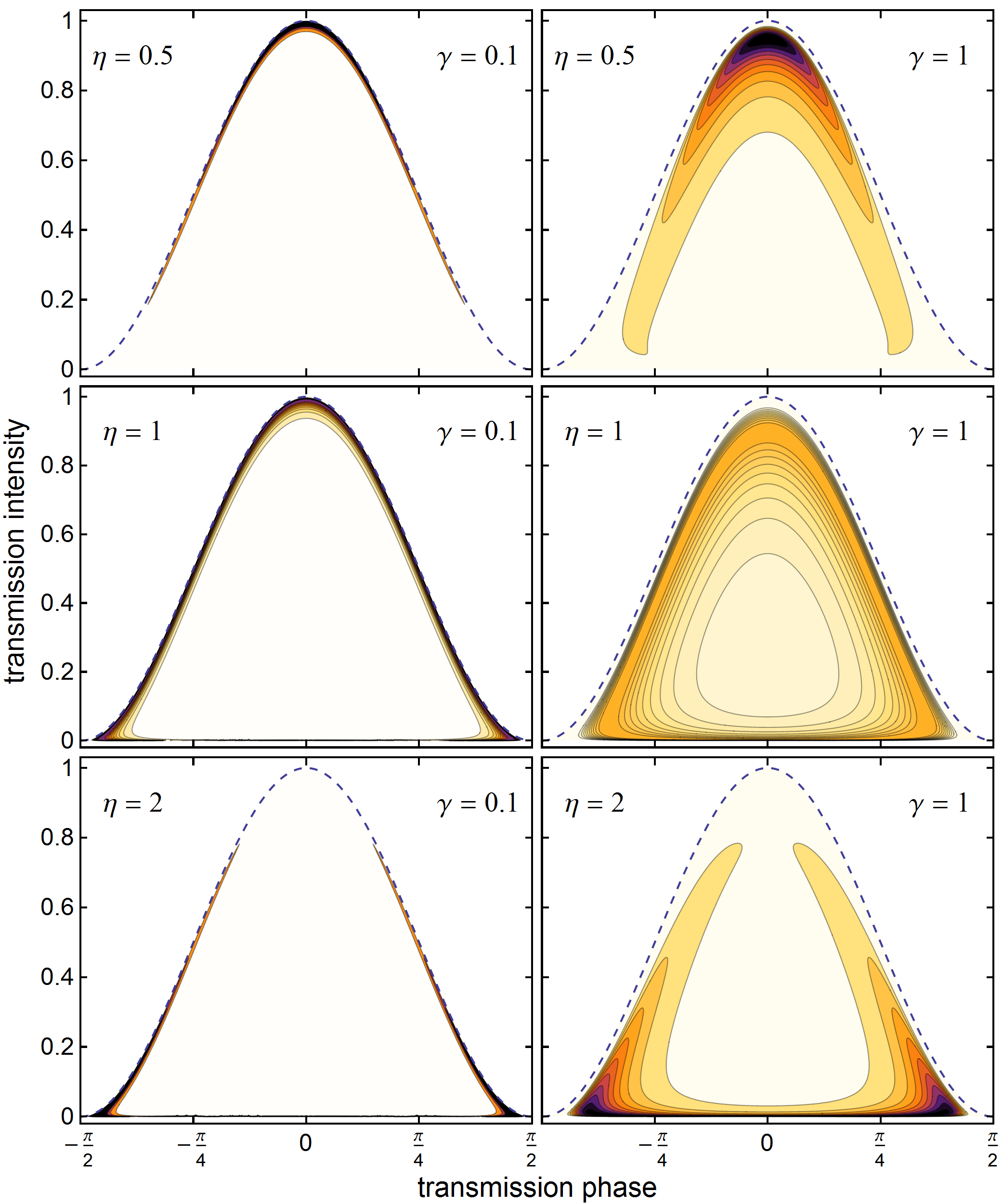}
  \caption{Contour plots of the joint distribution (\ref{Pjoint}) of the transmission intensity ($T$) and phase ($\theta$) at the background coupling $\eta=0.5,1,2$ (rows) and absorption rates $\gamma = 0.1,1$ (columns). The dashed line indicates the boundary $T=\cos^2\theta$ of the distribution support.
  \label{fig:T-phase} }
\end{figure}

In the limit of vanishing absorption, $\gamma=0$, one finds
\begin{equation}
  \mathcal{P}_{\gamma=0}(T,\theta) =  \delta(T - \cos^2\theta) \mathcal{P}_0(\theta)
\end{equation}
in agreement with the flux conservation constraints (\ref{T,R_gam0}). In the general case of finite absorption, the singularity of the joint distribution is removed, since $T$ and $\theta$ are no longer functions of each other. As was already mentioned, the function $P_0(x)$ gets exponentially suppressed at large $x\gg1$. As a result, the distribution at small $\gamma$ is mostly concentrated within a thin layer $\sim\gamma\ll1$ near the boundary $T=\cos^2\theta$.  When $\gamma$ is increased, the distribution starts exploring its whole support. Its weight is gradually moved from the central region around $T\sim1-2\eta$ at $\eta\ll1$ to a stripe around $T\sim\eta^{-2}$ at $\eta\gg1$. All these features are clearly seen on Fig.~\ref{fig:T-phase} showing the density plots of $\mathcal{P}(T,\theta)$ for various values of $\eta$ and $\gamma$.

It is worth discussing the statistical correlations between $T$ and $\theta$ in more detail. It is natural to expect that such correlations should go away, when absorption becomes strong. Making use of the exact limiting form $P_0(x)\approx\frac{\gamma}{4}\,e^{-\frac{\gamma}{4}(x-1)}$ at $\gamma\gg1$, we readily get the following asymptotic expression:
\begin{equation}\label{Pasym}
  \mathcal{P}^{\mathrm{(asym)}}_{\gamma\gg1}(T,\theta) =
  \frac{ \gamma\exp\bigl[ -\frac{\gamma(1+\eta)^2}{ 8\eta }
        \frac{T-2\aver{t}\sqrt{T}\cos\theta +\aver{t}^2 }{ \sqrt{T}(\cos\theta-\sqrt{T})}\bigr]
  }{ 16\pi T(\cos\theta-\sqrt{T})^2}\,,
\end{equation}
where $\aver{t}=(1+\eta)^{-1}$. This clearly shows that the correlations remain essential even at strong absorption.

Still, assuming very large $\gamma\gg1$, one can perform fluctuation analysis further. In such an extreme limit, one finds that $\sqrt{T}-\aver{t}$ and $\theta$ eventually become uncorrelated normal variables with the corresponding variances  $\sigma^2_T = \frac{4\eta^2}{\gamma(1+\eta)^4}$ and $\sigma^2_\theta=\frac{4\eta^2}{\gamma(1+\eta)^2}$. We note, however, that such a Gaussian approximation is very crude, because of the finite support of the exact distribution (\ref{Pjoint}). One can obtain a better approximation at strong absorption by studying the joint statistics of the real ($t_r$) and imaginary ($t_i$) parts of $t$ instead \cite{savi17b}. The two turn out to decorrelate faster than $T$ and $\theta$ when absorption grows. At $\gamma\gg1$, one finds that both $t_r-\aver{t}$ and $t_i$ become uncorrelated normal variables with the same variance $\sigma^2_T$. In such an approximation, finding the amplitude and phase distributions of $t_r=\sqrt{T}\cos\theta$ and $t_i=\sqrt{T}\sin\theta$ reduces to a classical problem studied by Rice~\cite{rice48} (see also \cite{yaco05}), yielding
\begin{equation}\label{Price}
  \mathcal{P}^{\mathrm{(rice)}}_{\gamma\gg1}(T,\theta) =
  \frac{1}{ 4\pi\sigma^2_T}e^{-(T-2\aver{t}\sqrt{T}\cos\theta+\aver{t}^2)/2\sigma^2_T} \,.
\end{equation}
The Rician approximation (\ref{Price}) resembles the exact asymptotic form (\ref{Pasym}) in its structure, but fails to properly take into account the boundaries of the distribution support. For that reason, it provides a reasonable approximation only at $\eta\approx1$, when the density is mostly concentrated in the centre, showing noticeable deviations otherwise, when the density gets concentrated near $T\sim1$ ($T\sim\eta^{-2}$) for small (large) $\eta$. Note that our asymptotic result (\ref{Pasym}) is free from such shortcomings, providing uniformly good approximation even at moderately large $\gamma$. We refer to Ref.~\cite{savi17b} for further discussion.

\begin{figure}[b]
  \centering
  \includegraphics[width=.925\linewidth]{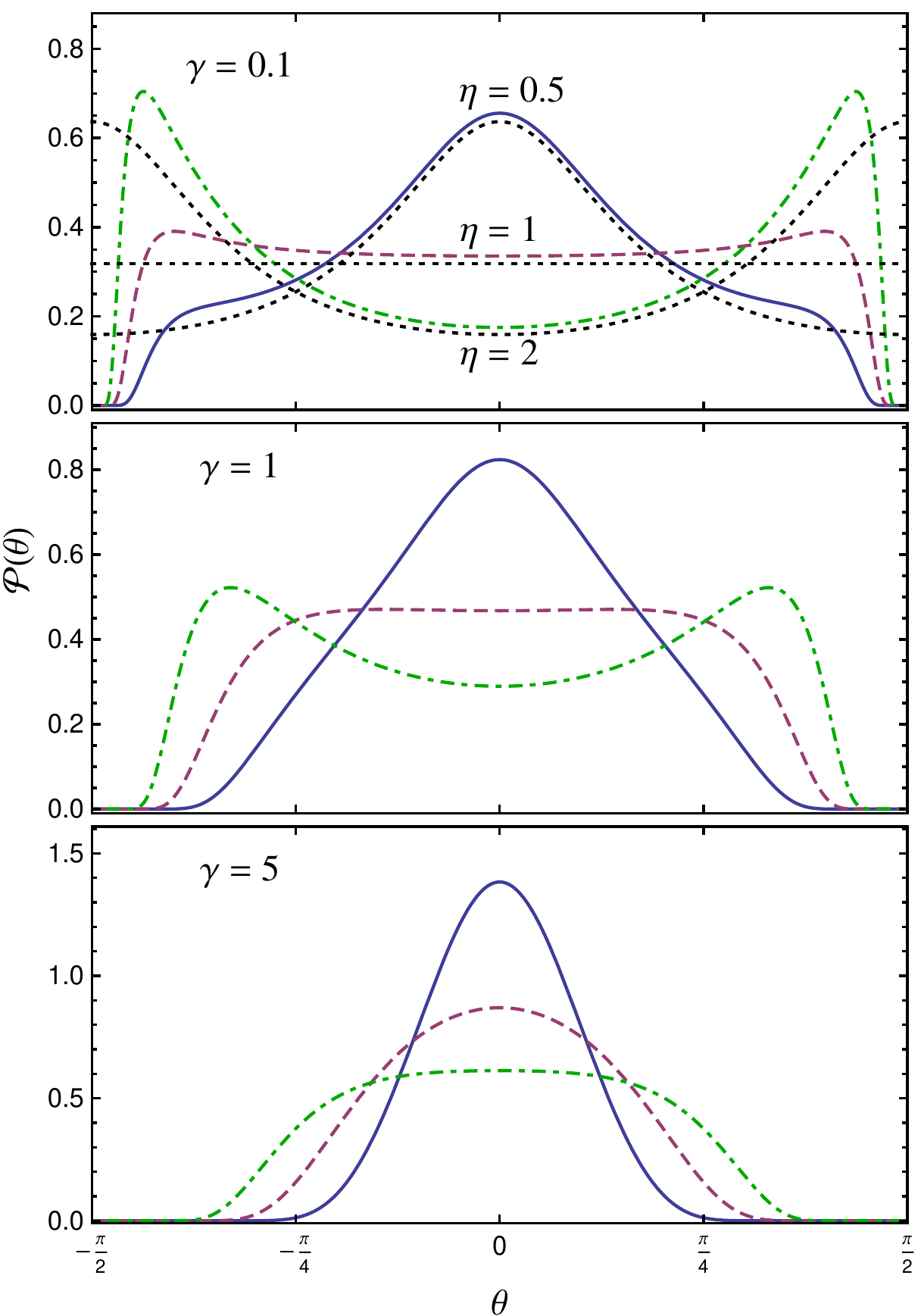}
  \caption{Distribution (\ref{Pphase}) of the transmission phase at the absorption rate $\gamma=0.1,1$ and $5$ (top, middle and bottom) and the background coupling $\eta=0.5$ (blue solid lines),  $1$ (pink dashed lines) and $2$ (green dot-dashed lines). The black dotted lines (top inset) show the zero absorption result (\ref{Pphase0}) for comparison.
  \label{fig:phase} }
\end{figure}

With the exact result (\ref{Pjoint}) in hand, one can now obtain both marginal and conditional pdf's by performing the relevant integrations. In particular, the distribution of the transmission intensity can be brought to the form (\ref{Ptr}) discussed above. The distribution $\mathcal{P}(\theta)$ of the transmission phase is obtained by integrating (\ref{Pjoint}) over $T$ and reads as follows \cite{savi17b}
\begin{equation}\label{Pphase}
  \mathcal{P}(\theta) =  \frac{\sec^2\theta}{2\pi} \int_{0}^{\infty}\frac{dp}{p^2}
   (1+p)P_0 [x(p,\theta)],
\end{equation}
where $\sec\theta=(\cos\theta)^{-1}$ and $x(p,\theta)$ is defined by
\begin{equation}\label{x}
  x(p,\theta) =\frac{(1+p)^2\sec^2\theta - 2p + \eta^2 - 1}{2\eta p}\,.
\end{equation}
With an explicit formula for $P_0$ found in \cite{savi05}, expression~(\ref{Pphase}) provides the exact result at arbitrary $\eta$ and $\gamma$.

Further analysis is possible in the limits of weak and strong  absorption, utilizing simpler asymptotic forms of $P_0$ as before. One finds the following approximation at small $\gamma$:
\begin{equation}\label{Pphase_weak}
  \mathcal{P}_{\gamma\ll1}(\theta)\approx\mathcal{P}_{0}(\theta)
     \Bigl[\mathrm{erfc}(\sqrt{\mu}) + 2\sqrt{\mu/\pi}e^{-\mu}\Bigr],
\end{equation}
where $\mu=\frac{\gamma}{4\eta}(\sec^2\theta-1+\eta)$ and $\mathrm{erfc}(z)=1-\mathrm{erf}(z)$ is the complementary error function. The bulk of distribution (\ref{Pphase_weak}) is essentially given by that at zero absorption, Eq.~(\ref{Pphase0}). The correction factor becomes crucial near the edges, where the exact distribution has an exponential cutoff $\sim e^{-(\gamma/4\eta)\sec^2\theta}$.

In the opposite case of strong absorption, $\gamma\gg1$, we find
\begin{equation}\label{Pphase_str}
  \mathcal{P}_{\gamma\gg1}(\theta)\approx \frac{\gamma\sec^2\theta}{4\pi}
  \Bigl[K_0(\xi) + \frac{\gamma\sec^2\theta}{4\eta\xi} K_1(\xi)\Bigr]e^{-\nu},
\end{equation}
with $\xi=\frac{\gamma}{4\eta}\sec\theta\sqrt{\sec^2\theta-1+\eta^2}$, $\nu=\frac{\gamma}{4\eta}(\sec^2\theta-1-\eta)$, and $K_{n}(z)$ being the modified Bessel function. In the limit of very large $\gamma$, the phase distribution tends to a Gaussian with zero mean and  the variance $\sigma^2_\theta$ provided above. Figure~\ref{fig:phase} shows the exact phase distribution for various $\eta$ and $\gamma$.

\subsection{Phase rigidity and mode complexness}\label{sec:phase}

Let us now discuss an application of the above results in the context of the so-called phase rigidity \cite{v.lang97}. It is a useful measure to quantify the influence of the environment resulting in complex-valued field patterns \cite{rott09}. In open chaotic billiards, e.g., such a complexness reveals itself in long-range correlations of the wave function intensity and current density \cite{brou03} that were studied experimentally \cite{kim05}. Following \cite{pnin96,lobk00a}, it is also convenient to characterise the above-mentioned complexness through a related parameter, the ratio $q^2$ of the squares of the imaginary and real parts of the relevant complex field. Such a $q$-factor was studied in microwave billiards \cite{bart05a}, where it can be linked with the presence of nonhomogeneous losses \cite{savi06b}. For weakly open chaotic systems, the RMT approach to the $q$-factor statistics was developed in \cite{poli09b}.

In the present case, it is natural to consider the phase rigidity of the transmission amplitude defined by
\begin{equation}\label{rho}
  \rho = \frac{t_r^2-t_i^2}{t_r^2+t_i^2} = \frac{1-q^2}{1+q^2} = \cos2\theta.
\end{equation}
Therefore, fluctuations of $\rho$ are directly induced by those of $\theta$. The corresponding distributions are related as follows
\begin{equation}\label{Prho}
  \mathcal{P}(\rho) = \frac{1}{\sqrt{1-\rho^2}}\mathcal{P}(\theta)|_{\sec^2\theta=2/(1+\rho)}\,,
\end{equation}
for $-1\leq\rho\leq1$, and similarly for $q^2=\frac{1-\rho}{1+\rho}$. Distribution (\ref{rho}) can be fully described using the results presented above.

In the limit of zero absorption, $\gamma=0$, Eq.~(\ref{Pphase0}) results in
\begin{equation}\label{Prho0}
  \mathcal{P}_0(\rho) = \frac{2}{\pi\sqrt{1-\rho^2}[\eta+\eta^{-1}+(\eta-\eta^{-1})\rho]}.
\end{equation}
This distribution develops a square root singularity $\sim\frac{1}{\sqrt{1-|\rho|}}$ at the edges and has the following symmetry under the involution $\eta\to\eta^{-1}$: $\mathcal{P}_0(\rho)|_\eta=\mathcal{P}_0(-\rho)|_{\eta^{-1}}$. It is interesting to note that the functional form (\ref{Pphase0}) already appeared earlier in a different context \cite{kanz96}, where it describes the distribution of the phase of the complex wave function induced by an external magnetic field. (Then $\eta$ is playing the role of the strength of that field.) As discussed there (see also \cite{lobk00a}), such a form arises from the assumption for the real and imaginary parts of the wave function to be uncorrelated Gaussian variables with different variances. For the transmission amplitude, however, such an assumption is simply not applicable, as $t_r$ and $t_i$ are deterministic functions of each other at zero absorption. Here, distribution (\ref{Pphase0}) appears by a different reason (which can actually be related to the so-called Poisson kernel).

At finite absorption, the correlations between $t_r$ and $t_i$ remain strong, decreasing gradually with the increase of $\gamma$. One finds that the phase rigidity distribution (\ref{Prho}) has an exponential cutoff $\sim\exp[-\frac{\gamma}{2\eta(1+\rho)}]$ as $\rho\to-1$, whereas the square root singularity at $\rho\to1$ remains unaffected. Of course, one can perform more detailed analysis of (\ref{Prho}) making use of the exact form of the phase distribution provided by Eq.~(\ref{Pphase}).

\subsection{Conclusions}\label{sec:summary}

We have presented a systematic study of fluctuations in resonance scattering induced by coupling the transmitting resonance to the chaotic background. Our approach combines the strength function formalism to account for the interaction with the background and the RMT modelling to mimic the chaotic nature of the latter. It enables us to obtain a number of the nonperturbative results for various statistics of the scattering observables. This includes the joint and marginal pdf's of the reflection and transmission intensities and phases that are derived in exact forms valid at arbitrary coupling to and losses in the background. The intensities and phases are found to develop strong and non-trivial statistical correlations, which remain essential even in the limit of strong absorption. In the latter case, we discuss the relevant approximations and their accuracy. In particular, a simple asymptotic expression (\ref{Pasym}) for the joint intensity-phase distribution has been obtained that, in contrast to the Rician distribution, provides good uniform approximation within the whole distribution support.

The obtained results can be used, e.g., to quantify the statistics of total losses or to study the phase rigidity (or mode complexness) in such a scattering situation. We note that it has now become possible to measure the full $S$ matrix, including the phases, in various microwave cavity experiments \cite{kuhl05,kuhl05a,hemm06,koeb10,diet10,yeh12,kuhl13,grad14}. In particular, exact analytical predictions for the statistics of diagonal \cite{fyod05,savi05} and off-diagonal \cite{kuma13,nock14} $S$ matrix elements were tested with high accuracy in such studies (see \cite{kuma17} for the most recent analysis). Therefore, we expect the results presented here to find further applications within a broader context of wave chaotic systems.

\subsection*{Acknowledgments}

I am grateful to the Organisers of the 8th Workshop on Quantum Chaos and Localisation Phenomena in Warsaw, Poland, May 19--21, 2017, where a part of this research was presented. I thank my collaborators of Ref.~\cite{savi17} for useful discussions of preliminary experimental results obtained with a reverberation chamber setup developed at INPHYNI, Nice.

%

\end{document}